\documentstyle[12pt]{article} 
\newcommand{\newsubsection}[1]{
\vspace{1cm}
\pagebreak[3]
\addtocounter{subsection}{1}
\addcontentsline{toc}{subsection}{\protect
\numberline{\arabic{section}.\arabic{subsection}}{#1}}
\noindent{ \it  #1}                 
\nopagebreak
\vspace{2mm}
\nopagebreak}
\newcommand{\newsection}[1]{
\vspace{15mm}
\pagebreak[3]
\addtocounter{section}{1}
\setcounter{equation}{0}
\setcounter{subsection}{0}
\setcounter{footnote}{0}
\addcontentsline{toc}{section}{\protect
\numberline{\arabic{section}}{{\rm #1}}}
\noindent{\sc \thesection.  #1}                 
\nopagebreak
\vspace{2mm}
\nopagebreak}
\newlength{\extraspace}
\newlength{\extraspaces}
\newcommand{\be}{\begin{equation}
\addtolength{\abovedisplayskip}{\extraspaces}
\addtolength{\belowdisplayskip}{\extraspaces}
\addtolength{\abovedisplayshortskip}{\extraspace}
\addtolength{\belowdisplayshortskip}{\extraspace}}
\newcommand{\ee}{\end{equation}}
\newcommand{\ba}{\begin{eqnarray}
\addtolength{\abovedisplayskip}{\extraspaces}
\addtolength{\belowdisplayskip}{\extraspaces}
\addtolength{\abovedisplayshortskip}{\extraspace}
\addtolength{\belowdisplayshortskip}{\extraspace}}
\newcommand{\ea}{\end{eqnarray}}
\newcommand{\nonu}{\nonumber \\[2mm]}
\newcommand{\is}{& \!\! = \!\! &}
\newcommand{\twomatrixd}[4]{{\left(\begin{array}{cc}
\displaystyle #1 & \displaystyle #2\\[2mm]
\displaystyle  #3  & \displaystyle #4 \end{array}\right)}}

\newcommand{\half}{{\textstyle{1\over 2}}}

\newcommand{\Z}{{\bf Z}}           
\newcommand{\C}{{\bf C}}  
            
\newcommand{\cF}{{\cal F }}            
\newcommand{\cZ}{{\cal Z }}            
            
\newcommand{\ra}{\rightarrow}

\renewcommand{\Im}{{\rm Im\,}}

\newcommand{\ext}{{\raisebox{.2ex}{$\textstyle \bigwedge$}}}

\input epsf
\newcounter{fignum}
\newcommand{\figuurnum}{\arabic{fignum}}
\newcommand{\figuur}[3]{
\addtocounter{fignum}{1}
\addcontentsline{lof}{figure}{\protect
\numberline{\arabic{section}.\arabic{fignum}}{#3}}
\hspace{-3mm}{\it fig.}\ \figuurnum.
\begin{figure}[t]\begin{center}
\leavevmode\hbox{\epsfxsize=#2 \epsffile{#1.eps}}\\[3mm]
\parbox{11cm}{\small \bf Fig.\ \figuurnum: \it #3}
\end{center} \end{figure}\hspace{-1.5mm}}

\newcommand{\Tr}{{\rm Tr}\,}

\newcommand{\qbold}{{\hbox{\boldmath $q$}}}
\newcommand{\qqbold}{{\hbox{\footnotesize \boldmath $q$}}}
\newcommand{\UU}{\sigma}%{\mbox{\footnotesize\it U}}%
\newcommand{\TT}{\rho}%{\mbox{\footnotesize\it T}}%
\newcommand{\VV}{\upsilon}%{\mbox{\footnotesize\it V}}%

\begin{document}
\addtolength{\baselineskip}{.5mm}

\begin{flushright}
July 1996\\ {\sc cern-th}/96-170\\ %{\sc }-????
\end{flushright}
\vspace{-.3cm}
\thispagestyle{empty}

\begin{center}
{\Large\sc{Counting Dyons in $N=4$ String Theory}}\\[13mm]

{\sc Robbert Dijkgraaf}\\[2.5mm]
{\it Department of Mathematics}\\
{\it University of Amsterdam, 1018 TV Amsterdam}\\[6mm]
{\sc Erik Verlinde}\\[2.5mm]
{\it TH-Division, CERN, CH-1211 Geneva 23}\\[.1mm]
and\\[.1mm]
{\it Institute for Theoretical Physics}\\
{\it Universtity of Utrecht, 3508 TA Utrecht}\\[4mm]
and \\[3mm]
{\sc  Herman Verlinde}\\[2.5mm]
{\it Institute for Theoretical Physics}\\
{\it University of Amsterdam, 1018 XE Amsterdam} \\[.1mm]
and\\[.1mm]
{\it Joseph Henry Laboratories}\\
{\it Princeton University, Princeton, NJ 08544}\\[18mm]
{\sc Abstract}
\end{center}

\noindent
We present a microscopic index formula for the degeneracy of dyons in
four-dimensional $N=4$ string theory. This counting formula is
manifestly symmetric under the duality group, and its asymptotic
growth reproduces the macroscopic Bekenstein-Hawking entropy. We give
a derivation of this result in terms of the type II five-brane
compactified on $K3$, by assuming that its fluctuations are described
by a closed string theory on its world-volume.  We find that the
degeneracies are given in terms of the denominator of a generalized
super Kac-Moody algebra. We also discuss the correspondence of this
result with the counting of D-brane states.

\vfill

\newpage

\newsection{Introduction}

In this paper we will study the dyonic spectrum of $N=4$ string theory
in 4 dimensions. This theory has two perturbative formulations: one in
terms of the toroidally compactified heterotic string, and a dual
formulation in terms of type II strings compactified on $K3\times
T^2$. To describe the set of charged states we will take the point of
view of the heterotic string, so that we have 28 electric charges
$q_e$ and 28 magnetic charges $q_m$ which both lie on the
$\Gamma^{22,6}$ lattice. This theory is conjectured 
\cite{font,sen,gauntlettharvey} to have as its
exact duality group \be SL(2,\Z)\times SO(22,6,\Z) \ee with $SL(2,\Z)$
being the electric-magnetic duality symmetry. The perturbative
heterotic string states carry only electric charge, so that states
with magnetic charge arise necessarily non-perturbatively as solitons.

Most accessible to computations is the spectrum of BPS states. These
states are annihilated by some subset of the 16 supersymmetry charges
and therefore many of their properties are protected in perturbation
theory. The purely electric states are special in the
sense that they preserve $1/2$ of the supercharges and their
degeneracies are easily determined, since they simply correspond to 
the heterotic string states that are in the
right-moving ground state. Hence a BPS  heterotic string state is
described by specifying the 28 charges $q_e \in \Gamma^{22,6}$ 
together with the occupation numbers $N^I_l$ ($I=1,\ldots,24$, $l>0$), 
subject to the level-matching condition 
\be
\label{levl}
\half q_e^2 + \sum_{l,I} l N^I_l = 1.
\ee
Here the subscript $l$ to $N^I_l$ denotes the world-sheet oscillation 
number of the coordinate field $X^I$, and $q_e^2$
is defined using the $SO(22,6)$ invariant inner product on the
$\Gamma^{22,6}$ lattice.
The number of such states is given by\footnote{Here and in the
subsequent we omit a factor of 16, and therefore count the number 
of 1/2 BPS multiplets rather than individual states.}
\be 
\label{hets}
d(q_e) = \oint \!d\UU\, {e^{i\pi \UU q_e^2}
\over \eta(\UU)^{24}}  
\ee 
where the `contour' integral over $\UU$ is from $0$ to $1$ and $\eta(\UU)$ is 
the Dedekind $\eta$-function.

The conjectured electric-magnetic $SL(2,\Z)$-duality predicts that
there should also exist a solitonic version of the heterotic string
that carries pure magnetic charge $q_m \in \Gamma^{22,6}$, and thus
that a similar formula counts the pure magnetically charged 1/2 BPS
states. The generic dyonic states, however, preserve only $1/4$ of the
supersymmetries and are more mysterious.  In the following we will
give a concrete proposal for the exact degeneracy of these dyonic BPS
states. To be more precise, we will present a formula for the number
of bosonic minus the number of fermionic BPS-multiplets for a given
electric and magnetic charge.  The basic idea is the following.  We
like to think of the dyonic states as some kind of bound state of an
electric heterotic string with a dual magnetic heterotic string.  The
formula that counts these states, should therefore be a suitable
generalization of (\ref{hets}), that will contain (\ref{hets}) as a
special subcase. In addition it has to satisfy other non-trivial
consistency checks, such as invariance under the full duality group,
while it also has to reproduce the correct asymptotic growth as
predicted from the macroscopic entropy formulas of extremal dyonic
black holes \cite{bh,cvetic}.

After presenting our formula and illustrating some of its features, we
will show that the complete duality invariant dyon spectrum has a
unified and natural interpretation in terms of strings on the
five-brane.  We will also discuss the correspondence with the counting
of BPS-states that follows from a D-brane description
\cite{polch,stromingervafa,blackholes,4dblackholes}.

\newsection{The degeneracy formula}

To write the proposed dyonic index formula, it will be convenient
to combine the electric and magnetic charge vectors $q_e$ and $q_m$
into a vector as follows 
\be
\qbold = \pmatrix {q_m\cr q_e} 
\ee 
Accordingly we introduce the $2\times 2$ matrix
\be 
\label{omega}
\Omega = \twomatrixd \TT \VV \VV \UU 
\ee 
generalizing the single modulus $\UU$ in (\ref{hets}). 
The degeneracy formula will then take the following form
\be
\label{degeneracy}
d(q_e,q_m) = \oint \!d\Omega\, {e^{i\pi \qqbold \cdot \Omega \cdot
\qqbold} \over {\Phi(\Omega)}} .
\ee 
The integrals over the moduli parameters $\UU, \TT$ and $\VV$ again all run
over the domain from $0$ to $1$, and impose level matching conditions
analogous to (\ref{levl}).

In the above suggestive form, it is natural to identify the matrix
$\Omega$ with the period matrix of a genus two Riemann surface. In the
result that we will obtain, the denominator $\Phi(\Omega)$ indeed will
turn out to be a genus two modular form. More precisely, it is the unique automorphic form of weight 10 of the modular group
$Sp(2,\Z)$ (represented by $4\times 4$ matrices) and can be
expressed as the squared product of all genus 2 theta
functions with even spin structure
\be
\label{phitheta}
\Phi(\Omega)= 2^{-12}\prod_{\alpha=even} \theta[\alpha](\Omega)^2.
\ee
$\Phi(\Omega)$ also happens to be equal to the denominator in the 
Weyl-Kac-Borcherds character formula for a generalized super Kac-Moody 
algebra, and as such it is a special case of the
automorphic forms constructed in \cite{borcherds}.
The $SL(2,\Z)$ duality transformations are identified with
the subgroup of $Sp(2,\Z)$ 
that leave the genus two modular form $\Phi(\Omega)$ invariant,
and thus the presented degeneracies are manifestly duality symmetric.
In addition, the formula satisfies a number of other non-trivial
consistency checks, that we will now discuss. Subsequently, we will 
turn to the derivation and more detailed description of this result.

To make our formula somewhat more explicit, we can expand the integrand as
a formal Fourier series expansion 
\be 
{1\over \Phi(\Omega)} = \sum_{k,l,m} D(k,l,m)
e^{-2\pi i (k\TT + l \UU + m \VV)} 
\ee 
with $k,l,m$ integers. The coefficients $D(k,l,m)$ are all 
integers, that via (\ref{degeneracy}) give us the degeneracy 
for a given electric and magnetic charge as
\be 
d(q_m,q_e)=D(\half q_m^2, \half q_e^2,q_e\!\cdot\! q_m) 
\ee 
Note that the normalization of (\ref{phitheta}) is chosen such that
$d(0,0)=1$.

\newsubsection{Correspondence with the heterotic string}

As we will explain in the following, the parameter $\VV$ couples to
the helicity $m$ of the dyonic states. The integral over $\VV$
thus projects on dyons with helicity equal to zero. However, instead of
integrating over $\VV$ we can also put it to fixed value, like $\VV=0$.
The resulting formula will then correspond to a helicity trace $(-1)^m$,
which will project out all 1/4 BPS representations, except for
the purely electric or magnetic 1/2 BPS representations. Hence, 
for $\VV \rightarrow 0$ the integrand in formula (\ref{degeneracy}) 
should match with the degeneracy formula (\ref{hets}) for the heterotic 
string BPS spectrum.

To make this correspondence explicit, we note that the function 
$1/\Phi(\Omega)$ is in fact identical to the chiral half of the
bosonic string partition function (or rather, that of 24 massless
scalars) on the corresponding genus two curve (see \cite{moretwo}).
This means in particular that, if we 
factorize the genus two surface parametrized by $\Omega$ 
into two separate genus one surfaces with moduli $\TT$ and $\UU$,
we indeed recover the separate $1/2$ BPS partition sums.  In this analogy
the electric and magnetic charges correspond to the loop momenta through 
the two handles, as in \figuur{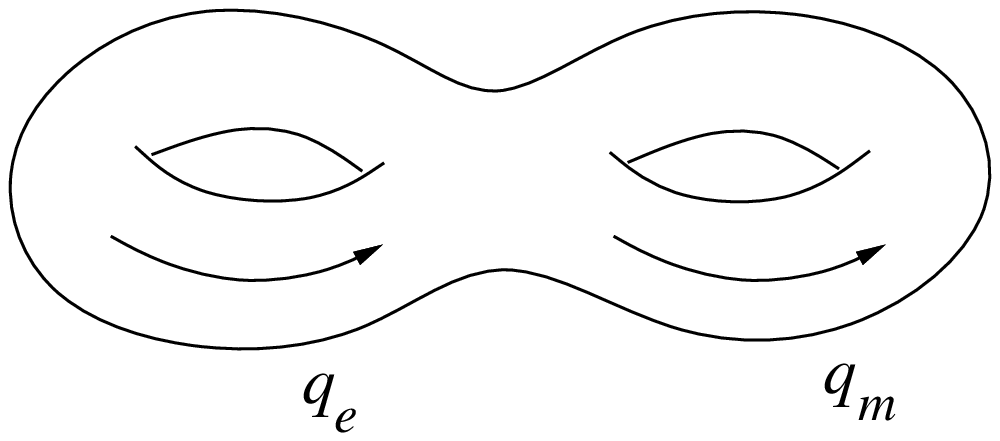}{6cm}{The dyon counting can be 
represented in terms of a genus two partition sum. The electric and magnetic 
charges of the dyon correspond to the loop momenta through the two 
handles of the Riemann surface.}

Concretely, this factorization corresponds to taking the limit
$\VV\rightarrow 0$, under which the function $1/\Phi(\Omega)$ reduces
to 
\be 
{e^{i\pi \qqbold \cdot \Omega \cdot \qqbold}\over \Phi(\Omega)}\ \
\longrightarrow \ \ {1\over \VV^2} \;
{e^{i\pi \TT q_m^2}  \over \eta(\TT)^{24}}\; 
{e^{i\pi \UU q_e^2} \over \eta(\UU)^{24}} 
\ee 
The diverging factor $1/\VV^2$ corresponds to the `tachyon pole,' and 
reflects the emergence of two non-compact translational zero-modes for
$\VV=0$. In the other factors we recover the separate electric and magnetic
heterotic string contributions.  The above correspondence is a
necessary boundary condition on any (candidate) dyonic degeneracy
formula.

\newsubsection{Asymptotic growth and black hole entropy}

As a further non-trivial boundary condition, the formula (\ref{degeneracy}) 
also matches the asymptotic behavior for large charges that one obtaines 
by comparing with the results for the macroscopic Bekenstein-Hawking 
entropy of extremal 4-dimensional black holes. This correspondence 
predicts that the statistical entropy of states with charge $(q_e,q_m)$ 
is given by \cite{cvetic}
\be
S = \pi \sqrt{q_e^2 q_m^2 - (q_e\!\cdot\! q_m)^2} .
\ee 
The asymptotic growth of the degeneracies $d(q_e,q_m)$ 
indeed agrees with this predicted entropy, as it can be shown that 
in the large charge limit 
\be 
\label{entropy}
d(q_e,q_m) \sim e^{\pi\sqrt{q_e^2q_m^2 - (q_e\cdot q_m)^2}} .  \ee A
formal derivation of this result is outlined in the Appendix, and
makes use of the zeroes of the function $\Phi$.  These zeroes lead to
poles in the integrand in (\ref{degeneracy}).  In our case the leading
contribution to the integral comes from the pole at $\TT\UU-\VV^2 +
\VV=0$%k\TT + l\UU + m\VV + n = 0$, with $m^2-4kl+4n=1$ 
(which is an appropriate $Sp(2,\Z)$-transform of the above described
`tachyon pole'). After evaluating the residue at this pole one can
perform the remaining integration by means of a saddle-point
approximation, and this immediately reproduces the above formula for
the asymptotic growth of $d(q_e,q_m)$. We will give an alternative
explanation of this result at the end of the next section, after we
have reinterpreted the degeneracy formula (\ref{degeneracy}) in terms
of strings living on a five-brane.

\newsection{Derivation from the quantum fivebrane}

It is known that the heterotic string arises as a soliton in type II
string theory that can be described as a five-brane wrapped around the
internal $K3$ manifold \cite{harveystrominger}.  $SL(2,\Z)$ symmetry
in the type II formulation is a consequence of $T$-duality, and thus
string-string duality implies that both the electric and magnetic
heterotic string can be considered as five-branes wrapped around
$K3$. Based on the above the degeneracy formula, we will argue in the
following that this correspondence can be naturally extended to all
dyonic states.

\newsubsection{Heterotic string from the five-brane}

The five-brane we will consider arises as a soliton in the
10-dimensional type II effective string theory \cite{soliton}. 
The long-wavelength
fields on its world-volume are given by an $N$ = (2,0) supermultiplet
consisting of a tensor with self-dual 3-index field strength $T_3$, 4
chiral fermions and 5 scalar fields. These fields represent the zero
modes of the 10-dimensional massless fields on the five-brane
background, which, via the standard adiabatic argument, are allowed to
vary along the five-brane world-volume.

The relation of the five-brane with the heterotic string can be
derived by dimensionally reducing the six-dimensional chiral field
theory on the world-volume to two dimensions on $K3$. In this way the
self-dual tensor field reduces to 19 left-moving and 3 right-moving
scalars. Together with the additional five scalars and fermions, this
gives exactly the world-sheet fields of the heterotic string. Moreover,
the string momentum lattice $\Gamma^{20,4}$ corresponds to the 
cohomology lattice of $K3$. In particular, the fluxes of the self-dual 
tensor field $T_3$ are identified with the electric charges
\be 
q_e^A
= \int_{S^1 \times \Sigma^A}T_3 
\ee 
where $S^1$ is one of the cycles
of $T^2$ and $\Sigma^A$ is a basis of two-cycles in $K3$.

In our setup, we would in fact 
like to take this correspondence one step further, by replacing the
standard world-sheet theory of the heterotic string by the
5+1-dimensional world-volume theory of the five-brane wrapped around
$K3$. Concretely this means that in counting the states of this
generalized heterotic string, we wish to take seriously the
possible fluctuations of the five-brane in the internal $K3$. At first
sight this would appear to imply a very drastic extension of the
world-sheet degrees of freedom, but it turns out that for purely
electrically charged BPS states nothing has
changed: when one imposes the BPS conditions one automatically finds
that the allowed fluctuations of the five-brane are the usual
left-moving heterotic string oscillations \cite{paper}. 
We will outline this calculation in a moment, after we have explained
our general set-up.

\newsubsection{Idea of the computation}

We would like to extend the above description to include dyonic states. 
As we will argue in the following, these dyonic states
will naturally arise as 1/4 BPS configurations of the five-brane,
provided we make one important new assumption. Namely, we will 
assume that the world volume description of the five-brane
is not described by a field theory, but rather by an appropriate 
closed string theory. This world-volume string theory must satisfy
the condition that the usual massless collective
fields should correspond to the ground states of this string, 
in direct analogy with the standard relation between 
the space-time effective field theory and the ground states of the 
type II string. In this sense, we wish to think about the
strings on the five-brane as essentially the original type II
string, but restricted to the world-volume.
The effective field theory of the string ground states is
sufficient to describe the long-wavelength fluctuations of the
five-brane, but at short distance we will also take into
account the other string modes, including winding configurations and
excitations.

Our aim is eventually to determine the degeneracy of
BPS states. We will do this by considering the 
partition function of the 5+1-dimensional world-volume string theory
with target space $K3\times T^2$.\footnote{This calculation is the direct 
analog of the conventional counting of string states by means of a 
conformal field theory partition function on a world-sheet of topology $T^2$.} 
In a light-cone gauge in which the transversal space
is taken to be the $K3$ manifold, the worldsheet theory on this
string is a $N=(4,4)$ superconformal field theory with $SU(2)_L\times
SU(2)_R$ current algebra \cite{letter,paper}. The eight supercharges on the 
world-sheet describe the unbroken part of the space-time supersymmetry. 
(The broken space-time supersymmetries give rise to zero-modes on the 
five-brane.)

The five-brane BPS partition function $\cal Z$ essentially counts the number 
of BPS states of this second quantized string theory. In computing this 
quantity, we will neglect string interactions. The justification for this 
should come from the fact 
that the BPS-condition in combination with the large number of 
supersymmetries severely restricts the possible interactions.
This expectation will indeed find concrete support in our analysis, as
it will turn out that for the relevant string partition sums
on the target space $K3\times T^2$, there are no higher loop correction 
to the free energy. Hence the fivebrane BPS partition sum can be computed by 
taking the exponent of the relevant one-loop free energy of the
string on the world-volume, provided one also adds an appropriate
zero-mode contribution 
\be
{\cal  Z} = \exp\Bigl[{\cal F}^{string}_{1-loop} + {\cal F}_{0-modes}\Bigr].
\ee  
The concrete form of this zero-mode contribution follows from
considering the free energy of the fluxes of the low-energy massless
field theory on the five-brane world-volume.
The partition sum ${\cal Z}$ will depend in particular on the moduli 
that parametrize the shape and size of world-volume manifold. 

To obtain the number of BPS states for given electric and magnetic charge, 
one finally needs to identify the corresponding contribution to the partition 
sum. In our formalism, this identification is automatically implemented 
via the integral over an appropriate set of world-volume moduli, which
act as Lagrange multipliers that impose `level-matching conditions'
analogous to the $L_0 - \overline{L}_0 = 0$ condition in string theory.
We will now explain and motivate this procedure in more detail for the 
1/2 BPS states, in which case we should recover the before-mentioned classical 
correspondence with the electric and magnetic heterotic 
string.

\newsubsection{Electric and magnetic fivebrane states}

To count the states that
respect half of the supersymmetries we have to restrict to the string
ground states. This turns the 6-dimensional string into a topological 
string  for which we can simply compute its one-loop free energy $\cal F$.
The one-loop computation is well-known \cite{dixon}. There are 24
ground states on $K3$, corresponding to the harmonic forms. We further
add the momenta and winding numbers in the $T^2$ direction, which take
value in the lattice $\Gamma^{2,2}$ parametrised by the
moduli $\TT,\UU$. Here $\TT$ parametrize the size and $B$-field on the
two-torus and $\UU$ parametrizes the complex structure. We then compute 
\cite{dixon}
\ba 
{\cal F} \is \half \int {d^2\tau\over \tau_2}
\sum_{(p_L,p_R)\in \Gamma^{2,2}} e^{i\pi(\tau p_L^2 - \overline\tau
p_R^2)} \cdot 24 \nonu \is 24 \log\left(\TT_2^{1/2}|\eta(\TT)|^2\right) +
24 \log\left(\UU_2^{1/2}|\eta(\UU)|^2\right) + cnst.  
\ea 
Here the first term represents the contribution of the winding 
modes of the strings that couple to the parameter $\TT$ and the second 
term the contribution of the momentum modes that couple to the 
parameter $\UU$. The occurrence of these two terms reflects a $\Z_2$ 
symmetry, which is an $R \ra 1/R$ symmetry applied to the world-volume 
of the five-brane. This transformation interchanges momentum and winding 
modes. So, even at the level of the string ground states, we encounter 
a pure stringy effect.

To obtain the 1/2 BPS partition function there are two
more steps to be taken. First, we have to take only the holomorphic
part of the above expression in $\TT$ and $\UU$, and secondly,
we need to include the zero-mode contributions. Specifically, 
we must replace the factor $\UU_2$ in ${\cal Z}=e^{\cal F}$  by the 
appropriate theta-function representing the summation of the fluxes 
over the lattice $\Gamma^{22,6}$, see \cite{erik}. 
Combined with the momentum contribution 
$1/\eta(\UU)^{24}$ this then gives the left-moving partition function of 
the heterotic string with modular parameter $\UU$. This is in accord with
the expected one-to-one correspondence between 1/2 BPS states of the 
five-brane and those of the heterotic string.

To restore the $\Z_2$ symmetry between $\TT$ and $\UU$, we also need
to replace the powers of $\TT_2$ by a lattice sum with a new set of
charges.  It seems natural to identify these new charges with the
magnetic charges $q_m \in \Gamma^{22,6}$. This identification finds
support from the heterotic-type II string duality, as follows.  The
electric-magnetic duality map that interchanges $q_e$ and $q_m$ acts
on the type IIA side as a $T$-duality on the $T^2$. Since the
five-brane we consider is wrapped around $K3$ and a one-cycle in
$T^2$, this transformation will act non-trivially on its world-volume
via an $R\ra 1/R$ transformation on the $S^1$. As we have just
discussed, this is a symmetry of the world-volume string theory that
interchanges the world-volume parameter $\TT$ and the parameter $\UU$.

Hence, we conclude that by representing the world-volume degrees of
freedom on the five-brane as strings, we are able to describe the
electrically charged heterotic string and the magnetically charged
heterotic string both in terms of one single fivebrane.

\newsubsection{Dyonic fivebrane states}

We now wish to consider five-brane states that carry
both electric and magnetic charge. These dyonic states necessarily
break $3/4$ of the 16 supercharges. Concretely this means that the
dyonic BPS five-branes carry more degrees of freedom than the purely
electric or magnetic configurations. 

As already noted, the 8 unbroken supercharges of the five-brane are
realized on the world-sheet of the string as the 4 left-moving and 4
right-moving generators of an $N=(4,4)$ superconformal algebra. So to
keep 4 unbroken charges we must restrict the string states to be in
the ground state of either the right-moving or the left-moving
sector. For definiteness, we will choose the right-moving one.
The index that counts these string states is given by the
elliptic genus of $K3$ \cite{ell}, which is defined as the weighted
trace in the RR-sector of the superconformal sigma-model
\be 
\chi_{\tau,z}(K3) =
\Tr (-1)^{F_L+F_R} e^{2\pi i(\tau (L_0-{c\over 24})+ zF_L)} 
\ee 
with $c=6$ for $K3$.  The fermion numbers $F_L$ and $F_R$ can be
identified with the zero-modes of the left-moving and right-moving
$U(1)\subset SU(2)$ current algebras.

As we restrict to right-moving ground states, $F_R$ only takes the
values $0,\pm 1$, while $F_L$ can a priori take all integer values and
gives us a conserved quantum number $m$. Since the corresponding
$SU(2)$-symmetry is part of the space-time Lorentz symmetry, this
conserved quantum number can be interpreted as the helicity in
space-time, {\it cf} \cite{vafawitten,spinningbh}. In the following we
will treat it on equal footing with the momentum and winding numbers
of the string around $T^2$. This turns the Narain lattice
$\Gamma^{2,2}$ of the two-torus into the lattice
$\Gamma^{3,2}$. Moreover, it allows us to extend the set of
world-brane parameters $\TT,\UU$ by another complex parameter $\VV$
that couples to the helicity quantum number $m$. Technically, $\VV$
has an interpretation as a Wilson loop on the world-volume that
parametrizes the $SU(2)_L$ bundle over $T^2$.  Together, $\TT$, $\UU$
and $\VV$ parametrize the lattice $\Gamma^{3,2}$, and transform in a
vector representation of the $SL(2,\Z)$ subgroup of the corresponding
$SO(3,2;\Z)$ $T$-duality group. (See Appendix.)

We now need to compute the following one-loop integral 
\be 
\cF = {\textstyle 1\over 2} \int {d^2\tau\over
\tau_2} \sum_{\mbox{\footnotesize $(p_L,p_R)\in\Gamma^{3,2}$} \atop
\mbox{\footnotesize $n \in 4\Z-\epsilon$}} e^{i\pi(\tau p_L^2-
\overline\tau p_R^2)} \cdot c(n) e^{i\pi \tau n/2}
\label{int}
\ee 
where $\epsilon=0,1$ depending on whether the helicity quantum number 
$m$ is even or odd, and
the coefficients $c(n)$ are defined by the expansion of the $K3$
elliptic genus as\footnote{Up to normalization this is the unique 
weak Jacobi form of index 1 and weight 0.}
\be 
\chi_{\tau,z}(K3) = \sum_{h\geq 0,\, m\in \Z}
c(4h-m^2)e^{2\pi i (h\tau+mz)} 
\label{K3-ell}
\ee 
It turns out that precisely the integral (\ref{int})
has been computed recently by T.\ Kawai in the context of $N=2$
heterotic string compactifications \cite{kawai} along the lines of
\cite{harveymoore}. The answer obtained in \cite{kawai} reads
\be 
\cF = -
\log\left[(\TT_2\UU_2- \VV_2^2)^{10} |\Phi(\TT,\UU,\VV)|^2\right] + const.
\ee 
where the holomorphic part has the following product representation 
\be 
\label{product}
\Phi(\TT,\UU,\VV)= e^{2\pi i(\TT +\UU +\VV)} \prod_{(k,l,m)> 0}\left(1 -e^{2\pi
i(k\TT+l\UU+m\VV)}\right)^{c(4kl-m^2)}.  
\ee 
Here $(k,l,m)> 0$ means that $k, l\geq 0$ and $m\in \Z$, $m<0$ for
$k=l=0$. (Note further that $c(n) = 0$ for $n<-1$).

Now, following the same logic as for the $1/2$ BPS states, we define
the partition sum for the dyonic $1/4$ BPS states by $\cZ = e^\cF$,
where again we have to include separately the zero-mode contributions
and restrict to the holomorphic sector. So the denominator of the
partition function is given by $\Phi(\TT,\UU,\VV)$ and represents the
contribution from multi-string states that describe the fluctuations
of the five-brane preserving the BPS condition. From the appearance of
the factor $\TT_2\UU_2 - \VV_2^2 = \det \Im \Omega$ it is clear that the
zero-modes must be included by a factor
\be 
\exp \Bigl[i \pi (q_m^2 \TT + q_e^2 \UU +
2q_e\cdot q_m \VV)\Bigr] 
\ee 
Combining all ingredients, and continuing the parallel with the
heterotic string discussion of the introduction, we finally arrive at the
announced result for the (bosonic minus fermionic) degeneracy formula 
\be
d(q_e,q_m) = \oint d\TT d\UU d\VV \, {e^{i \pi (q_m^2 \TT + q_e^2 \UU +
2q_e\cdot q_m \VV)} \over \Phi(\TT,\UU,\VV)}.
\ee

As was also noted in \cite{kawai}, it is natural to combine the moduli
$\TT$, $\UU$ and $\VV$ into a matrix $\Omega$ as in (\ref{omega}). In the
above form, the function $\Phi(\TT,\UU,\VV)$ is manifestly $SL(2,\Z)$
invariant, which suggests that it can be written as a modular
form in $\Omega$.  Indeed, a remarkable identity proved by Gritsenko
and Nikulin \cite{gritsenko} expresses the product representation
(\ref{product}) in terms of the product of even genus 2
theta-functions as in formula (\ref{phitheta}). The role of the 
automorphic form $\Phi$ in string theory was first recognized in \cite{mayr}.

\newsubsection{State counting}

The above single five-brane partition function has a concrete
interpretation as a trace of the multi-string quantum states of
a target space topology $K3\times S^1\times \bf R$. On such a target
space a string carries three additive quantum numbers: a winding
number $k$, a momentum $l$ along the $S^1$ and a helicity
$m$. Together the three integers $(k,l,m)$ form a vector on the
lattice $\Gamma^{2,1}$ with length-squared $m^2-4kl$, which is
invariant under the symmetry group
\be 
SO(2,1,\Z) \cong SL(2,\Z) 
\ee
We claim that this $SL(2,\Z)$ group should be identified with the
electric-magnetic duality group.  Level-matching on the string tells
us that the conformal dimension $h$ in the $K3$ sigma model must be
set equal to $h=kl$. If we look at the elliptic genus formula we see
that the number of single string states with given quantum numbers
$(k,l,m)$ is $|c(4kl-m^2)|$, and thus duality invariant.  Such a
string state represents a space-time boson or fermion depending on the
sign of $c(4kl-m^2)$. (The sign of $c(n)$ is $+1$ if $n\equiv 0$ (mod
4) and $-1$ if $n\equiv-1$ (mod 4), with the exception of $c(-1)=2$.)

A state in the multi-string Hilbert space is characterized by a set of
occupation numbers $N_{k,l,m}^I$ with
$I=1,\ldots,|c(4kl-m^2)|$. Hence, the degeneracy $d(q_e,q_m)$ counts
the number of independent multi-string states
$|\{N_{k,l,m}^I\}\rangle$ that satisfy the appropriate level-matching
conditions implemented via the integral over the moduli $\UU$, $\TT$ and
$\VV$. These take the form
\ba 
\half q_m^2 +  \sum_{k,l,m,I} k N^I_{k,l,m} \is 1 \nonu
\half q_e^2 +  \sum_{k,l,m,I} l N^I_{k,l,m} \is 1 \\[2mm]
q_e\!\cdot\!q_m + \sum_{k,l,m,I} m N^I_{k,l,m} \is 1\nonumber
\label{level}
\ea 
The first two relations are obvious generalisations of the
level-matching conditions of the heterotic string. In the case
of only electric or magnetic charges they indeed reproduce the usual
counting. The third line relates the sum of the helicities of the
individual strings to the quantity  $q_e \cdot q_m$. A suggestive
interpretation of this constraint is that  $q_e \cdot q_m$ represents
the contribution to the helicity as carried by the external 
electric-magnetic field of the dyon, when considered as made up from
a separate electric and  magnetic `point charge'. According to this 
interpretation, the third constraint
in (\ref{level}) essentially puts the total helicity of the dyonic
state equal to zero.

\newsection{Comparison with D-brane counting.}
  
Here we would like to indicate how this representation of the
BPS states is related to the more conventional construction in terms
of D-branes \cite{vafa6d,senD,stromingervafa,4dblackholes}. 
As we will argue, both the string representation as well
as the level matching relations have a natural interpretation from 
this point of view.

\newsubsection{Strings and D-brane intersections}

Let us consider a five-brane wrapped around the $K3$ compactification
manifold, so that its world-volume has the space-like topology of $K3
\times S^1$.  A pure five-brane only carries NS-NS charge. More
general BPS-configurations that also carry R-R charge can be obtained
by forming bound states with D-branes. For our purposes the most
convenient description of these bound states is given in the context
of type II B string theory. In this case we need to include 1-branes,
3-branes and 5-branes. When more of these D-branes are taken to lie
within the five-brane world volume, they will in general have
intersections \cite{senD,vafa6d}. The collection of BPS states that
corresponds to a given charge configuration are obtained by quantizing
the appropriate degrees of freedom of these intersections
\cite{stromingervafa}.

Since $K3$ has only non-trivial homology cycles of even dimension,
on the given topology of the five-brane of $K3\times S^1$, all the type IIB
D-branes will be wrapped around a corresponding element of $H_*(K3)$ 
times the $S^1$. Hence to a given D-brane configuration we can associate
a vector
\be
\label{homol}
q \in H_*(K3,\Z)
\ee
in the integral homology of $K3$. This vector becomes identified with
the charge vector
\be
q \in \Gamma^{20,4}
\ee
via the isomorphism of $H_*(K3,\Z) \cong \Gamma^{20,4}$, in which the
norm on $\Gamma^{20,4}$ is identified with the intersection form.  A
pair of 3-branes will typically intersect along a 1-dimensional string
wound one or more times around the $S^1$ direction, and similarly the
1-brane intersects with the 5-brane along a string around the
$S^1$. 

These strings formed by the D-brane intersections encode the internal
degrees of freedom of the five-brane in a completely similar fashion 
as the string introduced in section 3. In particular, its 
transversal oscillations are also described by an appropriate $N=4$
superconformal field theory with target space $K3$, where the
BPS-restriction again only allows for either left- or right-moving 
string oscillations.  The spectrum of these quantized `intersection-strings' 
will in particular include the excitations of the low-energy effective 
field theory on the five-brane world volume. The number of 
excited string states is again counted by means of the $K3$ eliptic genus.

It is evident from the above description that the total string winding 
number is given by the intersection pairing of the homology class (\ref{homol})
on $K3$ defined by the full D-brane configuration. Hence we see that
the D-brane representation gives a clear geometrical interpretation of 
the matching condition 
\be
\half q_m^2 +  \sum_{k,l,m,I} k N^I_{k,l,m} = 1.
\ee
Here we identified the vector in (\ref{homol}) with the magnetic charge
$q_m$.

In addition to the string winding number, the string oscillations will
also contribute to the total momentum along the $S^1$. This total
momentum is measured from the perspective of the five-brane world-volume,
and is therefore related but not equal to the internal KK momentum. 
The relation
between the two quantities is determined via a level matching condition 
equating this total string momentum to a zero-mode contribution.
As before, the form of the zero-mode contribution is determined by means 
of the free field theory of the massless five-brane modes, and thus depends
quadratically on the fluxes and KK-momenta  that gives rise to the electric 
charge $q_e$. This gives the second condition
\be
\half q_e^2 +  \sum_{k,l,m,I} l N^I_{k,l,m} = 1.
\ee
as the constraint of translation invariance on the world-volume
along the $S^1$ direction.

Finally, since the D-brane intersection strings carry a non-zero space-time 
helicity $m$, we can also introduce a third level matching condition
that equates the sum of the individual string contributions to the 
total helicity quantum number.

\newsubsection{$T$-duality}

We can learn more about the dyonic five-brane configuration
by considering the effect of the $T$-duality transformation that
interchanges the electric and magnetic charges. Concretely, let us
consider the configuration of a dyonic five-brane wound along the $A$-cycle
of the internal two-torus $T^2$, and act on it with the simultaneous 
$R \rightarrow 1/R$ transformation on both the $A$- and $B$-cycle. 
This map leaves the five-brane location invariant. However, when
applied to the intersection strings within the five-brane, it changes 
their winding direction to the $B$-cycle, while in addition it interchanges
the winding and momentum quantum numbers. It would appear therefore
that the new intersection strings lie perpendicular to the five-brane.

An alternative conclusion, however, is that a dyonic five-brane
should rather be thought of a pair of perpendicular five-branes,
one around the $A$-cycle and one around the $B$-cycle. Both five-branes
contain a collection of D-brane intersection strings, which however
are $T$-duals of each other in that the momentum quantum numbers of 
the $A$-cycle strings  are identified with the winding quantum numbers
of the $B$-cycle strings. In this picture, the total winding number of 
the intersection strings around, say, the $A$-cycle is measured by 
$q_m^2$, while the total string winding number around the $B$-cycle 
is $q_e^2$.

\newsubsection{A generating function}

It has been argued that the above multiple D-brane description leads to a
two-dimensional sigma model with target space the sum of symmetric
products of $K3$ manifolds \cite{stromingervafa}.  The 1/2 BPS states 
are in one-to-one correspondence with ground states of this model
and correspond to harmonic forms. The number of these ground states
follows from a direct application of the so-called orbifold formula, 
and is summarized in terms of the generating function (see \cite{vafawitten})
\be 
\sum_n  e^{2\pi i n\UU} \dim
H^*\! \left({(K3)^n\over S_n}\right) = \prod_{l>0} {1\over
(1-e^{2\pi i l \UU })}{}_{24} 
\label{H}
\ee
Our results now suggest a concrete generalization of this counting formula
to the $1/4$ BPS states, in which case one has to include the
transversal string oscillations.

As was first suggested in \cite{stromingervafa},
the counting of these states is naturally related to  the elliptic genus of 
the above-mentioned sigma model. Based on our degeneracy formula, we 
conjecture that these elliptic genera have a generating function
\footnote{Here in the product we removed, relative to (\ref{product}), 
the contribution of the states with $l=0$. The resulting expression is 
exactly the contribution of the non-degenerate orbits in the fundamental 
domain integral \cite{kawai}.}
\be 
\sum_n  e^{2\pi i n\UU} \chi_{\strut \! \TT,\VV}\!\! 
\left({(K3)^n\over S_n}\right)
= \prod_{k\geq0,l>0,m\in\Z}{1\over \left(1 -e^{2\pi
i(k\TT+l\UU+m\VV)}\right)}{}_{c(4kl-m^2)}
\label{ell}
\ee This conjectured identity suggests the following physical
interpretation.  The terms on the left-hand side count the
oscillations of a single long string on the multiple product of the
transversal target space $K3$.  The right-hand side of our formula
(\ref{ell}) on the other hand gives the usual second quantized
representation of a multiple string state on $K3$ in terms of a string
field theory Fock space.  The equality of both sides, if indeed true,
indicates that the single long string state on the symmetric product
of $K3$ provides an exact first quantized representation of the
appropriately symmetrized multiple string state on $K3$. Loosely
speaking, we equate the elliptic genus of the `second-quantized
manifold' to the second-quantized elliptic genus of a single copy of
the manifold. Based on this intuition, we expect that in the above
relation the manifold $K3$ can be replaced by an arbitrary manifold
$X$, if one uses the expansion coefficients $c(n)$ of the elliptic
genus of $X$.

We note further that the above formula satisfies a number of highly
non-trivial consistency checks. First, the $n=1$ term gives indeed the
answer (\ref{K3-ell}) for the $K3$ elliptic genus. Secondly, for
$\VV=0$ one recovers the ground state expression (\ref{H}). Thirdly,
the expression on the right-hand side of equation (\ref{ell}) has the
required automorphic properties. Precisely because of that reason this
particular expression has been considered before in the literature
\cite{gritsenko}. It has the important property that, when expanded in
powers of $e^{2\pi i \UU}$ (the so-called Fourier-Jacobi expansion)
the $n$-th coefficient is a weak Jacobi form of weight zero and index
$n$ for the modular group $SL(2,\Z)$ that acts on the modulus $\TT$
and $\VV$. (See \cite{eichler} for the theory of Jacobi forms.) The
expansion coefficients are therefore indeed candidates for elliptic
genera of Calabi-Yau manifolds of (complex) dimension $2n$, which
matches the expansion on the left-hand side. Note that the space of
weak Jacobi forms of weight zero and fixed index is
finite-dimensional. Given the fact that the Euler characters match,
this gives extra support to our conjecture.

Finally, we note that the asymptotic growth of the right-hand side, as
described in section 2, gives via the above equality a prediction for
the number of single string states on $(K3)^n/S_n$ at large
oscillation level. This prediction can be verified by standard
techniques.

\newsection{Concluding remarks}

Let us finally mention some applications and possible generalizations
of our result.

\newsubsection{Five-dimensional degeneracies}

One of the consequences of our analysis is a concrete refinement of
the results of Strominger and Vafa \cite{stromingervafa} on the
counting of microscopic black hole states in five dimensions. The
formula (\ref{ell}) gives an explicit counting formula of these states
and in particular allows one to analyze the asymptotic growth in more
directions by separately controling the total string winding, momentum
and helicity number. In five dimensions the level matching conditions
equate these quantum numbers to respectively $q^2$, the charge $Q_H$
of \cite{stromingervafa} and the five-dimensional spin $J$. Explicitely,
the degeneracy of these states is given by
\be
d(q,Q_H,J) = \oint d\TT d\UU d\VV \, {e^{i \pi (q^2 \TT + Q_H \UU +
2 J\VV)} \over \Phi'(\TT,\UU,\VV)}
\ee
with $\Phi'$ the denominator on the right-hand side of equation
(\ref{ell}). The asymptotics of $\Phi'$ are similar to those of $\Phi$
as analyzed in the Appendix.

\newsubsection{Generalizations}

It would be interesting to extend our results to more general
four-dimensional black hole states, by allowing for the possibility 
of non-zero angular momentum \cite{spinningbh} or states away 
from extremality. 

Spinning extremal black holes are in fact most naturally included in
our description, although for this we would need to give up manifest
$SL(2,\Z)$-duality. The most general spinning dyonic black hole
geometry is characterized by {\it four} independent macroscopic
quantities (namely the angular momentum $J$, the length-squared of the
charges $q_e^2$, $q_m^2$, and $q_e\cdot q_m$).  Our most general
states, on the other hand, carry at most three macroscopic quantum
numbers. In our presentation we eventually arrived at a duality
symmetric expression by equating the total `internal' helicity to the
`external' helicity $q_e\cdot q_m$. However, we can in principle omit
this constraint and simply identify the total internal helicity with
the macroscopic external angular momentum. This will give a degeneracy
formula consistent with the predicted entropy for spinning black holes
\cite{spinningbh} in the case that $q_e\cdot q_m = 0$.

Another important generalization is to include near extremal black
hole states \cite{near}. Within our framework, an obvious step is to
relax the BPS condition on the individual strings, and allow for
arbitrary left- and right-moving oscillations.  We expect that a naive
extension of our formalism in this regime would indeed produce an
asymptotic number of states in accordance with the four-dimensional
macroscopic entropy proportional to the mass squared. This is in
essence a consequence of the fact that the mass-shell relation $L_0 =
$(mass)${}^2$ is applied twice in our formalism: once on the string on
the world-volume, and once on the five-brane in space-time. By a
similar argument we also expect that the energy gap will also behave
in accordance with the thermodynamic prediction.  We should stress
however that the resulting description will very likely be of limited
validity, since on various crucial points we have made use of the BPS
condition. In particular it is not at all clear that the (effective)
string theory on the world-volume can still be treated as an
(approximately) free string theory.

\newsubsection{Algebraic structure}

Our results suggest that there exists an interesting
algebraic structure behind the $N=4$ dyon spectrum, related
to a generalized Kac-Moody (GKM) algebra. The particular
(super)algebra that underlies our degeneracy formula
is a so-called automorphic correction \cite{gritsenko}
of a hyperbolic Kac-Moody algebra with three simple roots
with Cartan matrix
\be
\pmatrix{2 & -2 &-2 \cr
-2 & 2 & -2 \cr
-2 & -2 & 2 }
\ee
(Note that the corresponding 
GKM has an infinite number of imaginary simple roots, 
roughly corresponding to the single string BPS states.) The fact that
the denominator of this GKM appears (in the denominator) of
our partition function, suggests that perhaps the full characters,
including the numerator, also play a role in the BPS spectrum.
Perhaps it is possible to realize the spectrum generating
algebra of the $N=4$ dyon BPS states in terms of this symmetry
algebra (cf. \cite{giveonporrati}). 
Generalized Kac-Moody algebras also recently appeared in the $N=2$
context \cite{harveymoore}, where it was suggested that the heterotic
string vertex operators representing the perturbative BPS states form
a GKM.

\vspace{15mm}

{\noindent \sc Acknowledgements}

We would like to thank S. Bais, R. Borcherds, S. Ferrara,
G. v.d. Geer, C. Hofman, E. Kiritsis, C. Kounnas, G. Moore,
D. Neumann, L. Susskind, C. Vafa and E. Witten for helpful
discussions.  This research is partly supported by a Pionier
Fellowship of NWO, a Fellowship of the Royal Dutch Academy of Sciences
(K.N.A.W.), the Packard Foundation and the A.P. Sloan Foundation.

\vspace{15mm}
\pagebreak[3]

\renewcommand{\thesection}{A}
\renewcommand{\thesubsection}{A.\arabic{subsection}}
\addtocounter{section}{1}
\setcounter{equation}{0}
\setcounter{subsection}{0}
\setcounter{footnote}{0}
\noindent {\sc Appendix}
\medskip

The asymptotic behavior of the degeneracy formula (\ref{degeneracy})
are related to the zeroes of the automorphic form $\Phi$ in the
integrand.  These zeroes are well understood using either standard
facts on genus two Riemann surfaces, the results of Borcherds
\cite{borcherds} or the representation in terms of a string one-loop
amplitude \cite{harveymoore,kawai} as in (\ref{int}). From the latter
point of view we understand that $\Phi(\TT,\UU,\VV)$ acquires a zero
or pole whenever the moduli $\TT$, $\UU$ and $\VV$ that parametrize
the Narain lattice $\Gamma^{3,2}$ allow chiral vertex operators, {\it
i.e.} elements $p=(p_L,p_R)\in \Gamma^{3,2}$ with $p_R=0$. In terms of
the integer lattice vector $p=(k,l,m,a,c)\in\Gamma^{3,2}\cong \Z^5$ we
have \cite{kawai}
\be
p_R^2 = {1\over 2Y}|k\TT + l\UU + m\VV + a(\TT\UU-\VV^2) + c|^2
\ee
with $Y=\TT_2\UU_2-\VV_2^2$, and 
\be
p^2=\half m^2 - 2kl + 2ac.
\ee
If we use the vector notation 
$y=(\TT,\UU,2\VV)$, with $-\half y^2=\TT\UU- \VV^2$, we see that the
zeroes and poles occur at the `rational quadratic divisors' 
of \cite{borcherds}
\be 
-\half a y^2 + b\cdot y + c=0
\label{divisor}
\ee 
where $b=(k,l,m)\in\Gamma^{2,1}$ with norm-squared $b^2=-2kl
+\half m^2$ and $a,c\in\Z$.  We can write equation (\ref{divisor})
also as $p\cdot v=0$, where we have introduced the 5-dimensional null
vector $v=(y;-\half y^2,1)$. Of course $|p\cdot v|^2 \sim p_R^2.$ In
terms of the coefficients $c(n)$, the order of the zero of $\Phi(y)$
at such a divisor is $\sum_{n>0}c(-2n^2p^2)$ with $p^2>0$.  In our
case the only possibility is $c(- 1)=2$, so we necessarily have
$p^2=1/2$ and the partition function $1/\Phi(y)$ has a second order
pole at this divisor.
 
In the theory of genus two curves or abelian surfaces, the equation
(\ref{divisor}) is well-known to describe (a component of) the
so-called Humbert surface $H_\Delta$ with discriminant $\Delta=2p^2$
\cite{vdgeer}. In our case we are interested in the surface
$H_1$. This is indeed known to be the zero-locus of the product of all
even theta-functions.

The `tachyon pole' at $\VV=0$ occurs for $b=(0,0,1)$ and $a=c=0$,
which describes the factorization on a product of two elliptic
curves. All the other poles are transformations of this one by
$Sp(2,\Z)\cong SO(3,2,\Z)$. This group acts linear on the vectors $v$
and $p$, which is another way to see that the divisors are of the form
(\ref{divisor}). The leading contribution in the Fourier coefficient
\be 
D(q)=\int d^3\!y \,{e^{2\pi i q \cdot y}\over \Phi(y)} 
\ee 
for large $q$ from the divisor (\ref{divisor}) is of the form $
D(q)\sim e^{2\pi \mu |q|} $ with $\mu^2=(b^2+2ac)/a^2=1/2a^2$.  So the
dominant term will be given by $a=1$. This gives indeed the black-hole
entropy (\ref{entropy}). The other coefficients are then constrainted
by $4c-4kl+m^2=1$, so in particular $m$ is odd, and $k,l$ can be
arbitrary. However, since we have the freedom to shift the variables
$\TT,\UU,\VV$ by integers, we can put $k,l,c=0$ and $m=1$. The divisor
is then given by $\TT\UU-\VV^2+\VV=0$.

Let us finally make a few remarks about the geometrical interpretation 
in terms of a genus two surface. Let
$A_1,A_2,B_1,B_2$ be a basis for the homology group $H_1(\Sigma)$ of a genus
two Riemann surface $\Sigma$. Now consider the quotient of $\ext^2
H_1$ by the relation 
\be
A_1\wedge B_1+A_2\wedge B_2=0
\ee
and call the resulting 5-dimensional lattice $L\cong\Gamma^{3,2}$.  
It naturally
carries a signature $(3,2)$ bilinear form, coming from the
intersection form on $\ext^2 H_1$. This makes explicit the isomorphism
$Sp(2,\Z) \cong SO(3,2,\Z)$.  As a basis of
$L$ we choose
\be
B_1\wedge A_2,\
B_2\wedge A_1,\
A_1\wedge B_1=B_2\wedge A_2,\
B_1\wedge B_2,\
A_1\wedge A_2.
\label{basis}
\ee
There is a natural  map $L \ra \C$ given in terms of the 
periods of the abelian 
differentials $\omega_1,\omega_2$
\be
C \wedge D \ra \oint_C \omega_1 \oint_D \omega_2 - \oint_D\omega_1 
\oint_C\omega_2
\ee
Under this map our basis (\ref{basis}) gives the `bi-periods'
\be
\TT,\  \UU,\ \VV,\ \TT\UU-\VV^2,\ 1.
\ee
It is clear that symplectic transformations will act linear on these
variables.  To any basis element of $L$ we can associate a
zero-homology cycle in the homotopy group $\pi_1(\Sigma)$ by
\be
C\wedge D \rightarrow [C,D]=CDC^{-1}D^{-1}
\ee
Note that $[A_1,B_1][A_2,B_2]\sim 1$ so that this map is well-defined.
It associates a bi-period to a vanishing cycle. Indeed, in this way
the divisor $\VV=0$ describes the usual pinching along the dividing
cycle $[A_1,B_1]=[B_2,A_2]$. Other divisors can be similarly 
interpreted.

\renewcommand{\Large}{\large}

\end{document}